# Efficient Image Reconstruction and Practical Decomposition for Dual-energy Computed Tomography

Lei Li(李磊), Ai-Long Cai(蔡爱龙), Lin-Yuan Wang(王林元),
Bin Yan†(闫镔), Han-Ming Zhang(张瀚铭), Zhi-Zhong Zheng(郑治中),
Wen-Kun Zhang(张文昆), Wan-Li Lu(路万里), Guo-En Hu(胡国恩)

National Digital Switching System Engineering & Technological Research Centre,
Zhengzhou 450002, Henan, P.R. China

**Abstract:** Dual-energy computed tomography (DECT) can be potentially applied in advanced imaging fields because of its capabilities in material decomposition. However, image reconstruction and decomposition under sparse view datasets are influenced by various factors, such as insufficient data, noise, and inconsistent observations. To obtain high-quality CT images and decompositions, this paper proposes an iterative image reconstruction algorithm and a practical image-domain decomposition method for DECT. The reconstruction algorithm is formulated as an optimization problem, which includes total variation regularization term and data fidelity term. The alternating direction method is utilized to design the corresponding algorithm, which exhibits faster convergence speed than that of existing approaches. Image-domain decomposition applies penalized least square (PLS) estimation on decomposing material mappings. PLS includes a linear combination term and the regularization term, which increases the smoothness of estimation images. The authors implement and evaluate the proposed joint method on real DECT projections and compare the method with typical and state-of-the-art reconstruction and decomposition methods. Experiments on the dataset of an anthropomorphic head phantom show that our methods are effective in terms of noise suppression and edge reservation without blurring the fine structures.

## 1 Introduction

The invention of the dual-energy imaging theory has brought increasing developments for dual-energy computed tomography (DECT) [1-4], which demonstrates its promising applications in material decomposition [5, 6] and energy-selective imaging[7]. Differentiating materials can be potentially applied in medical diagnosis [8-13], such as in the splitting of bones and tissues [14, 15], angiography[8], imaging guided radiology therapy[16]. DECT may also be applied in industrial design, manufacturing, nondestructive inspection, and 3D printing. In the diagnostic energy range, the linear attenuation coefficient of any material can be approximated as the weighted summation of two different actual or virtual materials, which are referred to as basis images. The decomposition images can be used as diagnosis references in comparison with standard basis.

Within the context of DECT image reconstruction, the processing methods of energy-windowed intensity data are classified in two broad categories [17]: pre-reconstruction and post-reconstruction. The majority of processing methods for multi-window data also grouped into these categories. In the pre-reconstruction category, decomposition is directly carried out on raw projections followed by image reconstructions. The decomposition of post-reconstruction is implemented on CT images via numerical methods. In DECT, the decomposed materials are mainly determined by the attenuation difference of the two scans, which is closely related to signal cancellation. However, direct decompositions severely suffer from the degradation of signal-to-noise ratios (SNR) on resultant images [7, 18-20]. Previous works focus on improving SNR, and these works can be mainly divided into two categories depending on their incorporation into the reconstructions. The first class of methods apply smoothing procedures to reduce noise, such as low-pass filtration methods [20-24]. Niu[15] proposed an iterative image-domain decomposition method for filtered back-projection (FBP) reconstruction of images. Niu's method is effective only for full-view datasets. The second class of methods incorporates the decomposition into the reconstruction process. A model-based iterative reconstruction algorithm is proposed to relieve noise presence, which combines the decomposition and reconstruction[25, 26]. A regularization term [25] which preserves the edges in the image, is often included in the formulation to suppress noise while maintaining spatial resolution.

* Supported by National Natural Science Foundation of China (61372172)
† E-mail: ybspace@hotmail.com;

Dong[14] proposed an approach that combines iterative reconstruction and image-domain decomposition. One of the disadvantages of this method is that the CT images are easily overblurred.

However, one generic problem of the current DECT is that the scanning X-ray dose is an unavoidable problem [18, 19], particularly for diagnosis. DECT scanners with fewer doses are safer. High-quality reconstruction images rely on sufficient projections with relatively high tube current. These factors are inevitable in dose introduction, and dose reduction is crucial. The radiology dose in DECT imaging can be reduced by acquiring fewer projections with sparse views or using low tube current. However, under the condition of low tube current, the signal decomposition process is unstable and suffers from severe noise boost in the resulting images on the basis of experimental results and findings in the literature. We declare that low tube current is not considered in this paper. A hybrid scheme for reconstructing high-quality CT images on both full views and sparse views, as well as decomposing CT images into basis mappings that are implemented in an image domain, is proposed in this work.

Compressed sensing[27, 28] (CS) reconstruction algorithms, including iterative reconstruction using total-variation (TV) regularization[29], can be potentially used in recovering stable signals with superior SNR from noisy projection data. When the views of observation data are sparse or few, DECT imaging is not usually employed. This approach is adopted because image reconstruction cannot be implemented using conventional methods, such as FBP [30] or algebraic reconstruction approaches [31, 32]. State-of-the-art methods do not focus on the decomposition of basis mappings under such situations. Intermediate CT images of different energy levels are crucial in DECT imaging. Efficient approaches should provide high-quality images for linear attenuation coefficients and basis mappings.

The alternating direction method (ADM) [33-35], which has important applications [36-40] in CT reconstruction, is applied to address the convex problem. ADM applies variable-splitting method, which has simple derivations and stable convergence properties [41] and effectively addresses large-scale imaging problems. This paper proposes a general approach for both sparse and full-view datasets in DECT imaging, including CT image reconstruction and image-domain decomposition. An optimization scheme is established to describe the reconstructions from dual energy data.

The outline of this paper is as follows: Section 2 mainly describes the proposed method, including image reconstruction and decomposition. In Section 3, the experiments with real DECT dataset are used to compare the proposed method and existing approaches. The discussion and conclusion are presented in Section 4.

## 2 Method

### 2.1 Imaging model in DECT

In DECT imaging, the projection data are usually acquired by two different energy levels, i.e., high and low kVps. In our method, image reconstructions are first carried out, which are followed by image-domain decomposition. As shown in Figure 1(a), the methods and experiments presented in this paper are based on the cone-beam CT, which mainly consists of X-ray source, flat detector and the corresponding mechanical gantry system. The simplified geometry of the scanning system is described in Figure 1(b) which characterizes some essential geometry terms, e.g., distance of source to the rotation axis (SOD), distance of source to the detector (SDD), source orbit (red solid circle) and field of view (FOV, depicted by gray area in Figure 1(b) ).

The modeled system serves as the basis of image reconstruction. When the projection dataset is acquired under sparse views, discrete-to-discrete (DD) imaging model is frequently used, in which high and low kVps projection data can be presented by a linear system expressed as

$$\mathbf{A}\mathbf{x}_{h/l} = \mathbf{m}_{h/l}, \quad (1)$$

where $\mathbf{x}_{h/l} \in \mathbb{R}_+^N$ is the vector form of the image of linear attenuation coefficient at high ($\mathbf{x}_h$) or low ($\mathbf{x}_l$) energy level. The attenuation coefficients are usually nonnegative real numbers, and $\mathbb{R}_+^N$ is utilized to denote the $N$-dimensional nonnegative real vector space. Matrix $\mathbf{A} \in \mathbb{R}_+^{M \times N}$ is the cone–beam system matrix, and its elements are generated using the intersection length of the ray and each voxel in $\mathbf{x}_{h/l}$ by Siddon's model [42]. In this equation, the system matrix is determined according to the hardware settings, which are characterized by some critical parameters, i.e., SDD, SOD, scanning views and positions, and detector parameters. Projection dataset $\mathbf{m}_{h/l}$ is also known by observation. Attenuation coefficient $\mathbf{x}_{h/l}$ at different kVps is unknown which needs to be computed or reconstructed.

Image reconstruction aims to find the actual $\mathbf{x}_{h/l}$ according to the system matrix and observation projections. However, finding $\mathbf{x}_{h/l}$ by using an intuitive approach of converting system matrix $\mathbf{A}$ into its inverse is difficult. This is because of the following reasons: When the projection data are acquired in sparse or few views, the linear equation (1) is severely underdetermined, which means that $\mathbf{A}$ has a low rank and a very large condition number, which leads the computation is very unstable and easily affected by noise even at a low level. Moreover, in imaging fields, dimension $\mathbf{A}$ is usually very high, and its inverse is computationally expensive.

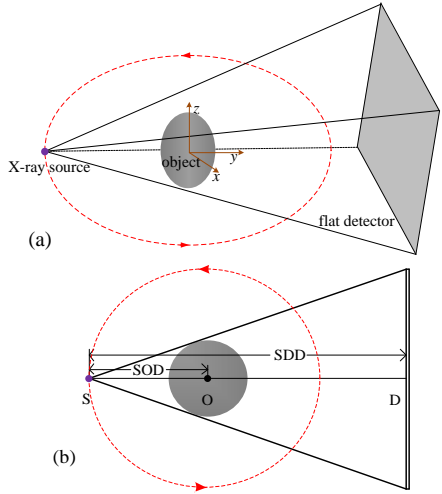

Figure 1 A simple sketch of the cone-beam CT (left) and the simplified geometry (right).

## 2.2 Alternating direction TV minimization image reconstruction

Finding attenuation coefficients $\mathbf{x}_{h/l}$ from a severely underdetermined system is the central task of DECT image reconstruction. In this work, the authors propose the TV-based image reconstruction model and its corresponding algorithm. The TV model was proposed and successfully applied in image denoising, in-painting, and deblurring. With the development of CS theory, TV becomes a frequently used sparse transform since that many images are composed of piecewise continuous or piecewise constant patches, particularly in CT imaging fields. CS theory states that many linear discrete (or digital) systems and signals can be exactly reconstructed or recovered from underdetermined observations with a high probability if the targeted signal is sparse or has some form of sparse transform.

In this work, we apply TV as the image regularization term in our reconstruction model. For a 2D image $\mathbf{x}$, which is the TV norm of $\mathbf{x}$ is expressed as

$$\|\mathbf{x}\|_{\mathrm{TV}} = \sum_{i,j} |\nabla \mathbf{x}_{i,j}| \\
= \sum_{i,j} \sqrt{\left(\mathbf{x}_{i,j} - \mathbf{x}_{i-1,j}\right)^2 + \left(\mathbf{x}_{i,j} - \mathbf{x}_{i,j-1}\right)^2}, \quad (2)$$

where $i, j$ denotes the index of $\mathbf{x}$ in two different directions. When $\mathbf{x}$ is composed of piecewise constant patches, $\nabla \mathbf{x}_{i,j}$ is highly sparse with very few nonzero entities only on the edges of $\mathbf{x}$. Thus, $\|\mathbf{x}\|_{\mathrm{TV}}$ is a small scalar and leads TV norm to be a frequently used regularization term.

A straightforward combination on $\mathbf{x}_{h/l}$ is carried out in the DECT image reconstruction model, in which two images of different kVps can be reconstructed simultaneously and can be expressed as

$$\mathbf{Wf} = \mathbf{g}, \\
\mathbf{W} = \begin{pmatrix} \mathbf{A} & \mathbf{0} \\ \mathbf{0} & \mathbf{A} \end{pmatrix}, \mathbf{f} = \begin{pmatrix} \beta_1 \mathbf{x}_h \\ \beta_2 \mathbf{x}_l \end{pmatrix}, \mathbf{g} = \begin{pmatrix} \beta_1 \mathbf{m}_h \\ \beta_2 \mathbf{m}_l \end{pmatrix}, \quad (3)$$

where $\beta_1$ and $\beta_2$ are positive scalars that are used to balance the noise level in the projection dataset and select the values that will be analyzed later in this paper. In this work, we apply TV norm on combined image $\mathbf{f}$. The reconstruction is modeled as an optimization problem where the object function is composed of a TV regularization term and data fidelity term:

$$\tilde{\mathbf{f}} = \arg\min_{\mathbf{f}} \|\nabla \mathbf{f}\| + \frac{\mu}{2} \|\mathbf{Wf} - \mathbf{g}\|_2^2, \quad (4)$$

where $\mu$ is a scalar used to balance the data fidelity and the TV regularization term. In developing an algorithm to solve the proposed optimization, auxiliary variable $\mathbf{z} = \nabla \mathbf{f}$ is introduced and used, which converts Equation (4) into the following form:

$$\tilde{\mathbf{f}} = \arg\min_{\mathbf{f}} \|\mathbf{z}\| + \frac{\mu}{2} \|\mathbf{Wf} - \mathbf{g}\|_2^2, \quad (5)$$

subject to $\mathbf{z} = \nabla \mathbf{f}$.

Two main variables exist in Equation (5), i.e., $\mathbf{f}$ and $\mathbf{z}$, and the structure of (5) indicate that it is proper and efficient to apply the alternating direction method (ADM). In the framework of ADM, the optimization problem is solved by transforming the form in Equation (5) into its augmented Lagrangian function, which can be depicted as

$$\left(\tilde{\mathbf{f}}, \tilde{\mathbf{z}}\right) = \arg\min_{\mathbf{f},\mathbf{z}} \|\mathbf{z}\| - \boldsymbol{\lambda}^{\mathrm{T}}(\nabla \mathbf{f} - \mathbf{z}) \\
+ \frac{\rho}{2} \|\nabla \mathbf{f} - \mathbf{z}\|_2^2 + \frac{\mu}{2} \|\mathbf{Wf} - \mathbf{g}\|_2^2, \quad (6)$$

where $\boldsymbol{\lambda} \in \mathbb{R}^N$ is the multiplier, and $\rho$ is a scalar that controls the TV stength. The optimization of (6) can be split into two sub-problems according to each independent variable (i.e., $\mathbf{f}$ and $\mathbf{z}$). ADM employs an iterative approach to reach the convergence solution of $(\tilde{\mathbf{f}}, \tilde{\mathbf{z}})$. The following approach is a most popular form:

$$\begin{cases} \mathbf{z}^{k+1} \leftarrow \arg\min_{\mathbf{z}} \|\mathbf{z}\| - \langle \boldsymbol{\lambda}, \nabla \mathbf{f}^k - \mathbf{z} \rangle \\ \qquad\qquad + \frac{\rho}{2} \|\nabla \mathbf{f}^k - \mathbf{z}\|_2^2, \\ \mathbf{f}^{k+1} \leftarrow \arg\min_{\mathbf{f}} \frac{\mu}{2} \|\mathbf{Wf} - \mathbf{g}\|_2^2 \\ \qquad\qquad + \frac{\rho}{2} \|\nabla \mathbf{f} - \mathbf{z}^{k+1}\|_2^2 - \langle \boldsymbol{\lambda}, \nabla \mathbf{f} \rangle, \\ \boldsymbol{\lambda}^{k+1} \leftarrow \boldsymbol{\lambda}^k - \xi(\nabla \mathbf{f}^{k+1} - \mathbf{z}^{k+1}), \end{cases} \quad (7)$$

where the operation "$B \leftarrow A$" means assigning the value of $A$ to $B$. The formula in (7) describes a single iteration loop, and ADM works in an iterative fashion.

The solution to the sub-problem of $\mathbf{z}$ in (7) has an efficient and analytical expression as:

$$\mathbf{z}^* = \max(\|\nabla \mathbf{f}^k - \lambda^k/\rho\| - 1/\rho) \cdot \text{sign}(\nabla \mathbf{f}^k - \lambda^k/\rho). \quad (8)$$

The solution to the sub-problem of $\mathbf{f}$ can be obtained using available methods, and one of the most efficient approach is the method developed by Li [43], which uses a one-step-gradient-descent method. The step size in one-step-gradient-descent is chosen using the adaptive Barziai–Borwein (BB)-like method. A simplified form of this method can be written as

$$\mathbf{f}^{k+1} \leftarrow \mathbf{f}^k - \alpha_{BB}^k \cdot G_{\mathbf{f}}^k, \quad (9)$$

where $G_{\mathbf{f}}^k = \mu \mathbf{W}^T (\mathbf{W}\mathbf{f} - \mathbf{g}) + \rho \nabla^T (\nabla \mathbf{f} - \mathbf{z}^{k+1}) - \nabla^T \lambda^k$ is the gradient with respect to $\mathbf{f}$ in k-th iteration loop, and the step size $\alpha_{BB}$ is set by BB like method. The DECT image reconstruction procedure is as follows:

$$\begin{cases} \mathbf{z}^{k+1} \leftarrow \max(\|\nabla \mathbf{f}^k - \frac{\lambda^k}{\rho}\| - 1/\rho) \cdot \text{sign}(\nabla \mathbf{f}^k - \frac{\lambda^k}{\rho}), \\ \mathbf{f}^{k+1} \leftarrow \mathbf{f}^k - \alpha_{BB}^k \cdot G_{\mathbf{f}}^k, \\ \text{update } G_{\mathbf{f}}^{k+1}, \\ \text{update } \alpha_{BB}^{k+1}, \\ \lambda^{k+1} \leftarrow \lambda^k - \xi(\nabla \mathbf{f}^{k+1} - \mathbf{z}^{k+1}). \end{cases} \quad (10)$$

### 2.3 Penalized least square image domain decomposition

DECT image reconstruction provides the attenuation coefficient of the high- and low-energy levels. By utilizing the reconstruction images at two different x-ray energies, DECT methods decompose the measured data and generate images of two basis materials. In the theory of image-domain material decomposition, the linear attenuation coefficient of each pixel in the CT image is approximated by the linear combination of the pixel values in the images of basis materials. In this paper, we assume that the approximation is accurate. The relationship of the material decomposition $\mathbf{u} = (\mathbf{u}_1; \mathbf{u}_2) \in \mathbb{R}^{2N}$ and the CT linear attenuation coefficient $\mathbf{x} = (\mathbf{x}_h; \mathbf{x}_l)$ can be written as

$$\mathbf{x} = \mathbf{D}\mathbf{u}, \quad (11)$$

where $\mathbf{D}$ is the material decomposition matrix with a dimension of 2N-by-2N where N is the total number of pixels in each CT image. $(\mathbf{u}_1(i), \mathbf{u}_2(i))$ is the pixel pair in the two images of the basis materials, which are the normalized densities of the base materials and are unitless. $\mathbf{D}$ is defined in this paper as

$$\mathbf{D} = \begin{pmatrix} \eta_{1H} I & \eta_{2H} I \\ \eta_{1L} I & \eta_{2L} I \end{pmatrix}, \quad (12)$$

where $I$ is the identity matrix with the dimension of N-by-N. $\mathbf{D}$ is a very sparse matrix, which can be computationally efficient in multiplying a vector. The subscripts H/L indicate the high/low energy spectrum, and the subscripts 1/2 denote the two materials bases. In (12), $\eta_{ij}$ is the linear attenuation coefficients of material $i$ (i=1 or 2) in mm$^{-1}$ under the spectrum of $j$ (j=H or L).

Direct decomposition using (12) generates the images of basis materials with severely degraded SNRs, compared with those of the raw CT images. This result occurs because dual energy ratio $\eta_{iH}/\eta_{iL}$ is not significantly different on the two basis materials in the diagnostic x-ray energy range, thereby leading to a large condition number on matrix $\mathbf{D}$. Thus, the decomposition is sensitive to noise on raw CT images.

This paper applies penalized least square estimation in decomposing the material mappings. The algorithm is written as

$$\tilde{\mathbf{u}} = \arg\min_{\mathbf{u}} \|\mathbf{u} - \mathbf{D}^{-1}\mathbf{x}\|_2^2 + \gamma R(\mathbf{u}), \quad (13)$$

where $R(\mathbf{u})$ is the regularization term to enhance the smoothness of $\mathbf{u}$, and $\gamma$ is used to adjust smooth strength. In this work, the specific form of $R(\mathbf{u})$ is set as the expression in [15]. To present a comprehensive study, we include the expression herein as

$$R(\mathbf{u}) = \frac{1}{2} \sum_i \sum_{k \in N_i} e_{ik} (\mathbf{u}(i) - \mathbf{u}(k))^2, \quad (14)$$

where $N_i$ is the set of four or six neighbors of the $i$-th pixel in the 2-D or 3-D image. Weights $e_{ik}$ is the edge detection weight, which is small if either $i$ or $k$ is the index of an edge pixel in the image and one if otherwise. Weights $e_{ik}$ can be set using existing edge detection operators, such as Canny method and Prewitt method. The objective function in (14) is a quadratic form and is convex. The optimal solution can be obtained from its first-order derivative as:

$$(2I + \gamma \nabla R)\mathbf{u} = 2\mathbf{D}^{-1}\mathbf{x}. \quad (15)$$

Equation (15) can be solved using many iterative approaches, such as standard gradient methods and their modified version. In this paper, preconditioned conjugate gradient (PCG) method is utilized to accelerate the convergence.

### 2.4 Implementations and evaluations

Some practical problems must be discussed in terms of specific implementation. The first issue is the selection of the values of the parameters. Three groups of parameters exist in the proposed reconstruction algorithm, i.e., $(\beta_1, \beta_2)$, $\rho$, and $\mu$. As mentioned earlier, $(\beta_1, \beta_2)$ is used to balance the noise level in the observation projections. The noise level of low kVps projections is higher than those in high kVps projections, thereby indicating that $\beta_1 < \beta_2$ when setting the values of $(\beta_1, \beta_2)$. $(\beta_1, \beta_2)$ can be computed by calculating and comparing the standard deviation of a small piecewise region in the projections. In this work, the ratio of $\beta_2 / \beta_1$ is empirically set within $(0.2, 1)$. The values of $\rho$ and $\mu$ are utilized to balance the regularization term and the data fidelity term in (6). The ratio of $\rho/\mu$ is crucial, and both values of $\rho$ and $\mu$ are relatively broad. A suggested range for the two values is $2^2$–$2^{10}$, and $\mu$ should be increased appropriately when the projection data are reduced. Parameter $\gamma$ in the decomposition algorithm should be chosen according to the SNR in the reconstruction images. This parameter can be approximately calculated using a small piecewise constant region of interest (ROI). In this work, $\gamma$ is relatively small, and the reasonable choice is within $(10^{-3}, 10^{-6})$.

The implementations of the proposed methods (10) and (15) are mainly written in MATLAB code. To speed up the computation of the forward and backward projection, the graphics processing unit (GPU, NVIDIA GeForce GTX 570) is utilized to accelerate the evaluations by coding CUDA C (NVIDIA, Santa Clara, CA) to apply the massive parallel computational capability of the GPU hardware. The fast forward and backward projection algorithms in this work are developed by Gao [44]. The CUDA C code interacts with the MATLAB code by the mex interface provided by MATLAB.

We mainly consider three algorithms in carrying out the comparisons in this paper. The first algorithm is the FBP plus the direct decomposition by applying numerical inverse. The second algorithm is proposed by Dong [14], which combined iterative reconstruction and image-domain decomposition for DECT by TV regularization. The third method is proposed by Niu [15], which directly applies the image-domain decomposition on FBP reconstruction. Niu's method involves least square estimation with smoothness regularization. The estimated variance-covariance matrix of the decomposed images as the penalty weight in the least square term exhibits great potential in noise suppression. The latter two algorithms in this paper are implemented according to authors' instructions of [14] and [15].

## 3 Real DECT data Experiments and Results

The experimental data are taken from a CBCT system. An anthropomorphic head phantom is utilized as the scanning object. Two groups of projection, which have the same geometry settings, are acquired at X-ray tube energies of 75 kVp and 125 kVp. Both groups have 655 projections over $2\pi$. The distance of X-ray source to the iso-center is 1000 mm, and the distance of the center of the detector to the iso-center is 500 mm. The pixel size of the detector is 0.388 mm. In our experiments, only a center slice of the cone–beam ray is used to test all the algorithms. Thus, the detector is actually a 1D discrete and finite line with 1024 bins. All the reconstruction and the decomposed images are composed of matrices with dimension of 512 x 512 with each pixel with a size of 0.5 x 0.5 mm$^2$.

In this work, the comparisons are carried out under two seminars, i.e., full views with all the 655 views over 360° of projections and sparse views with only 328 views over 180°. When the relative changes for the reconstruction images are less than the machine precision, the reconstructions stop. The run speed of the proposed algorithm under full view is approximately 0.52 seconds per iteration and 0.35 seconds for sparse view. The maximum iteration number for each reconstruction algorithm is set to 10000. In our implementations for the proposed method, several hundreds of iterations are sufficient to achieve a relatively fine image quality. The runtime of the proposed method is roughly 65 seconds (around 125 iteration loops) and 50 seconds (approximately 143 iteration loops) for full and sparse views, respectively. Dong's methods require approximately 1000 or more iteration numbers and take around 0.48 and 0.29 seconds per loop for each seminar. The PCG time consumption of the proposed PLS decomposition is approximately 20 seconds with a maximum iteration number of 500.

Comparing the different algorithms with real CT data is difficult because the background truth is theoretically inexistent. However, the FBP reconstructions with full view data are always used for clinical usage, the reconstruction are not true images because of the significant inconsistencies. In this work, we apply the FBP reconstructions, under full views and sparse views, to offer intuitive impressions of the CT images of the real projections.

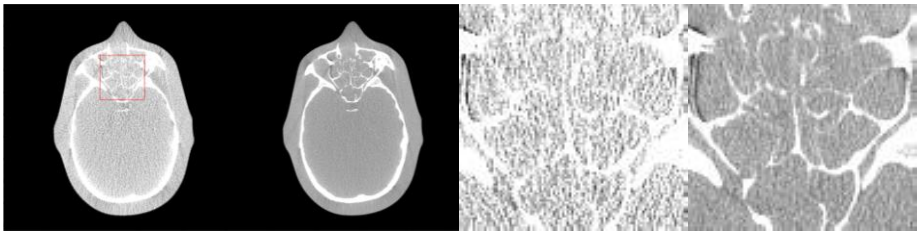

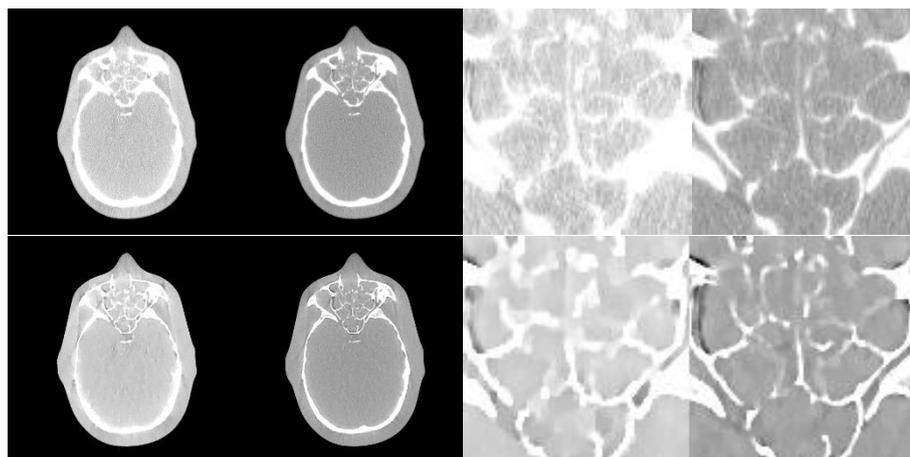

Figure 2: The reconstruction of each algorithm under full views of projections: From the top row to the bottom row, the results are generated by FBP, Dong's method and the proposed method, respectively. The first and the second column in the left are the reconstructions of low and high kVps. The third and the fourth column are the zoom-in displays of the region of the tiny structures in low and high energy levels, and the region is depicted by the red rectangle top-left image. The display window is set to [-500 HU 500 HU].

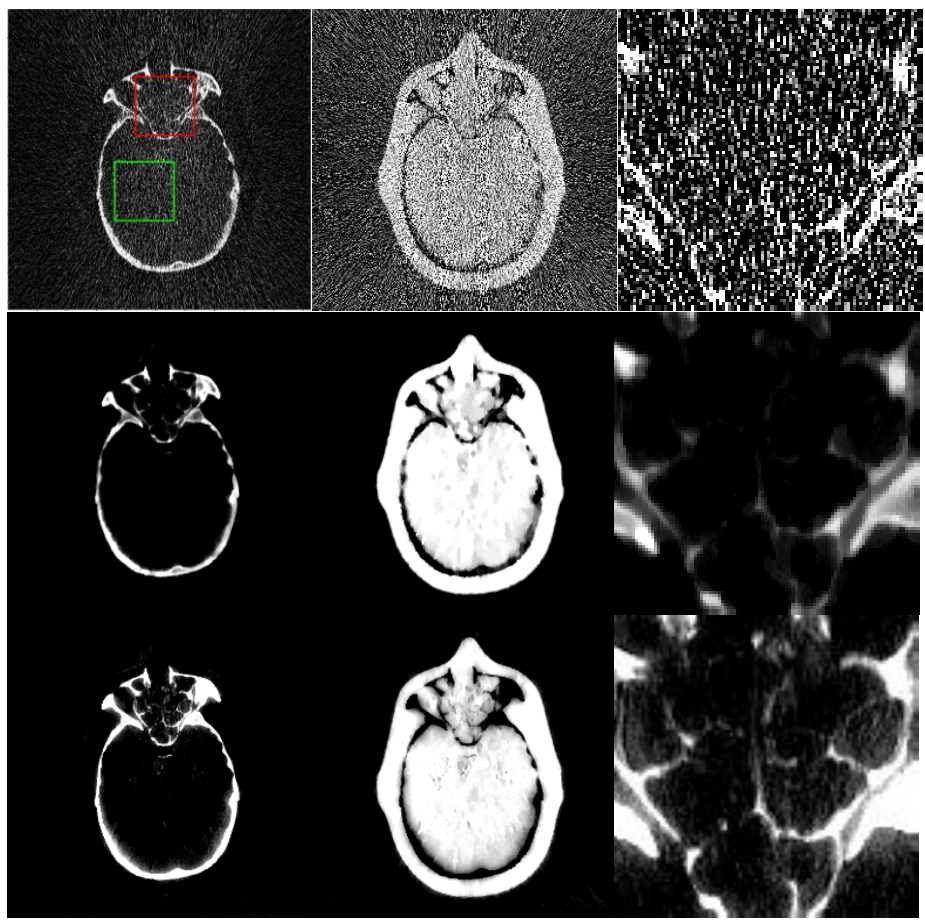

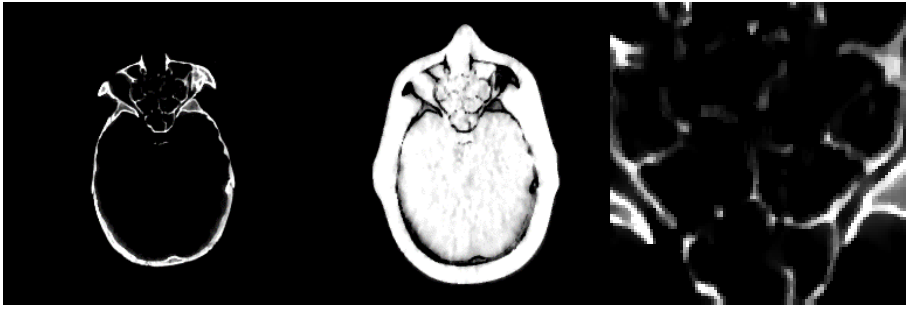

Figure 3: The decomposition comparison between each method under full views of projections: From the top to the bottom row, the results are generated by direct decomposition using numerical inversion for FBP reconstructions, Dong's method, Niu's method, and the proposed method. From the left column to the right column, the images are decomposed results of bone area, tissue area, and the local zoom-in displays of tiny structures in the decomposed bone area which are depicted by the rectangle in the top-left image. The display window is set to [0.1 1.0].

The details and the noise level in the resulting images can be a reasonable evaluation standard. The image details stand for the tiny and complicated structures in the scanning objects. For clinical applications, only the vision of doctors with extensive experiences and knowledge of anatomy can serve as a gold standard. However, in mathematical descriptions, the definition and evaluation of the details in the images are nontrivial. In this paper, we compare only the reservation of edges in the images. The ROIs of sinus structures of the head phantom are chosen to compare the edges. This region is depicted by the red rectangle in Figure 1. Noise level is evaluated by calculating the mean value and the standard deviation (SD) of a piecewise constant ROI. The second ROI is depicted by the green rectangle in Figure 3.

The results of the reconstructions with full view data, as well as the numerical results, are listed in Figures 2 and 3 and Table 1. The reconstructions of FBP, Dong's methods, and the proposed methods are shown in Figure 2. To clearly compare these images by visualization, they are shown in a relatively narrow display window of [-500HU 500HU]. The reconstructions of the proposed methods indicate the potential of these approaches in edge reservation, unlike that of FBP and the method proposed by Dong. The local zoom-in displays of a small ROI in the right portion of Figure 2 also validate this property. The reconstructions of FBP and Dong's method exhibit noise, particularly for low-energy level reconstructions in the zoom-in displays. Figure 3 shows the decomposed images of each method. The direct decompositions of the FBP reconstructions using numerical inversions suffer from severe noise, which causes the tiny structures to disappear within the noise. All the results of Dong's method, Niu's method, and the proposed methods show most parts of the tiny structures in the ROI. The proposed method suppresses the noise and maintains almost all these structures. The means and SDs of the pixel values inside the second ROI are summarized in Table 1.

The results of sparse views are listed in Figures 4 and 5 and Table 2. For sparse view data, FBP reconstruction exhibits severe noise and streak artifacts. All the images of both energy levels are not appropriate for diagnosis. The results of Dong's method and the proposed ones are also affected by data insufficiencies. The qualities of these reconstructions are degraded in different senses. A comparison indicates that the proposed method may exhibit a slight advantage in noise suppression and edge reservation, as shown in Figure 4. The decomposition images are presented in Figure 5, and all the images do not have the same quality as those in Figure 3. The performance of the proposed method in noise suppression and edge reservation is slightly better. The typical numerical results are listed in Table 2, which shows that the proposed method reconstructs more accurate images and provides better decompositions.

Table 1: The means and SDs of the pixel values inside the ROIs in linear attenuation images and the decomposed images depicted by the green rectangle in Figure 3, under full views data.

| Algorithms | linear attenuation images | | Decomposed images | |
|---|---|---|---|---|
|  | Low kVps | High kVps | "Bone" area | "Tissue" area |
| FBP and direct decomposition | 0.0244±0.0030 | 0.0214±0.0011 | -0.0156±0.5486 | 1.0397±1.2596 |
| Dong's method | 0.0244±8.147e-4 | 0.0214±6.2310e-4 | -0.0146±0.0358 | 1.0369±0.0747 |
| Niu's method | 0.0244±0.0030 | 0.0214±0.0011 | -0.0129±0.0460 | 1.0331±0.0624 |
| The proposed method | 0.0244±6.496e-4 | 0.0214±5.605e-4 | -0.0164±0.0303 | 1.0414±0.0659 |

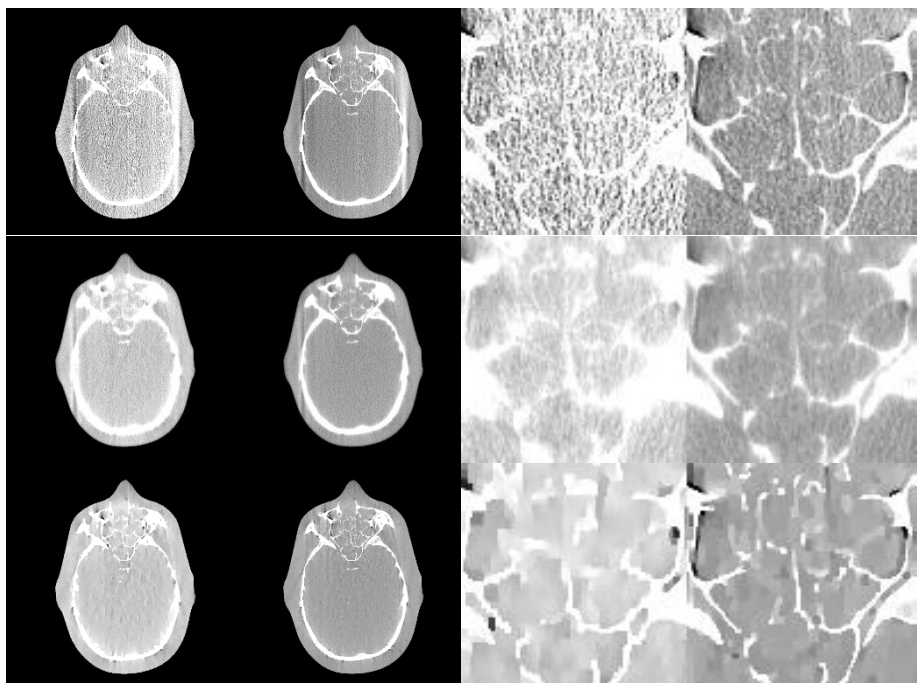

Figure 4: The reconstruction of each algorithm under sparse views of projections: Similar to Figure 2, from the top row to the bottom row, the results are generated by FBP, Dong's method and the proposed method, respectively. The first and the second column in the left are the reconstructions of low and high kVps. The third and the fourth column are the zoom-in displays of the region of the tiny structures in low and high energy levels, and the region is depicted the same with that in Figure 2. The display window is set to [-500 HU 500 HU].

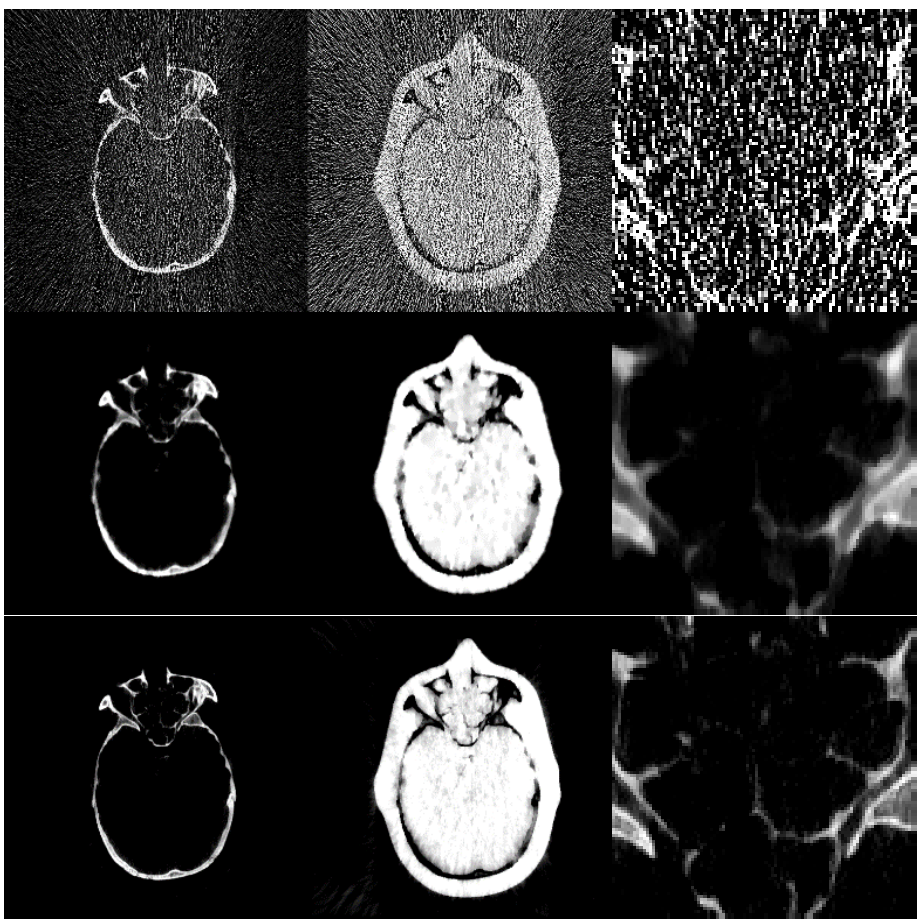

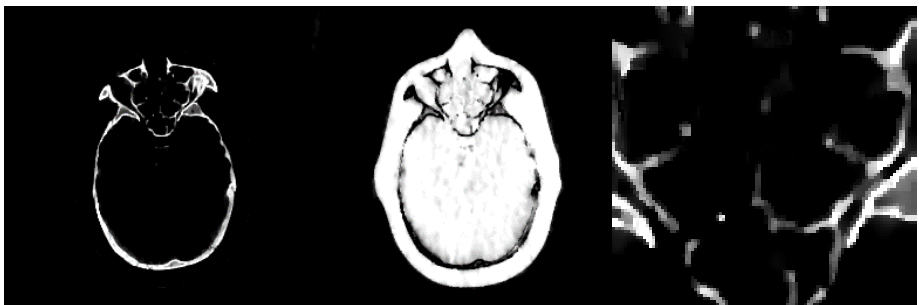

Figure 5: The decomposition comparison between each method under sparse views of projections: Similar to Figure 3, from the top to the bottom row, the results are generated by direct decomposition using numerical inversion for FBP reconstructions, Dong's method, Niu's method, and the proposed method. From the left column to the right column, the images are, respectively, decomposed results of bone area, tissue area, and the local zoom-in displays of tiny structures in the decomposed bone area the same with that in Figure 3. The display window is set to [0.1 1.0].

Table 2: The means and SDs of the pixel values inside the same ROIs depicted by the green rectangle in Figure 3, for sparse view data.

| Algorithms | linear attenuation images | | Decomposed images | |
|---|---|---|---|---|
| | Low kVps | High kVps | "Bone" area | "Tissue" area |
| FBP and direct decomposition | 0.0236±0.0036 | 0.0208±0.0013 | -0.0214±0.6656 | 1.0215±1.5270 |
| Dong's method | 0.0245±8.9116e-4 | 0.0215±6.5889e-4 | -0.0143±0.0496 | 1.0403±0.1058 |
| Niu's method | 0.0236±0.0036 | 0.0208±0.0013 | -0.0180±0.0510 | 1.0276±0.0816 |
| The proposed method | 0.0243±5.7966e-4 | 0.0215±4.0116e-4 | -0.0163±0.0319 | 1.0431±0.0742 |

## 4 Discussion and Conclusion

In DECT imaging, noise has different magnitudes for two different scanning energy levels. Noise causes the numerical inversion to be unstable and invalid for industrial or clinical applications. Developing practical reconstruction and decomposition methods for both full and sparse views is important. In this work, we propose a hybrid approach for DECT image reconstruction and image-domain decomposition. The method includes TV image reconstruction algorithm and PLS image-domain decomposition. The TV-based algorithm is developed by the ADM scheme, which is frequently utilized in optimization fields, particularly for sparse signal recovery with insufficient observation data. Moreover, TV regularization can be potentially used in noise suppression in reconstructing linear attenuations. The image-domain decomposition is established based on the PLS noise suppression model. A TV-like regularization term, which is inspired by Niu's method, is utilized. The PCG method is applied to solve the model.

This work applies the ADM framework to optimize the DECT reconstruction model, and a notable property of ADM over conventional algorithm is its fast convergence speed. The convergence property is also validated by the time consumption results of the proposed method. Dong's method needs more time than the proposed approaches to produce comparable image quality. For image decomposition parts, a simple PLS model is considered without the need to compute the variance-covariance matrix of reconstruction images. The high accuracy estimation of the variance-covariance matrix is nontrivial, which requires substantial data observations and sufficient statistical analyses prior to noise distribution. The experiments indicate that the proposed method provides comparable or slightly better image qualities on both seminars compared with that of Niu's method.

Future work is necessary to optimize the performance of the proposed method. Moreover, the improvement of regularization term is needed to achieve better noise suppression while maintaining the tiny and detailed structures in the objects. Extended tests and applications on industrial material identifications or real clinical usage are also highly significant.

## Acknowledgement

This work is supported by the National Natural Science Foundation of China (No. 61372172). We also want to thank the anonymous referees for giving us useful comments and suggestions to improve this paper.

## References

[1] Johnson TRC, Krauß B, Sedlmair M, et al., European radiology, **6**(14):1510-1517 (2007)

[2] Marin D, Boll DT, Mileto A, et al., Radiology, **271**(2):327-342 (2014)

[3] Muşturay K, Aykut A, Diagnostic and Interventional Radiology, **17**(3):181-194 (2010)

[4] Johnson TR, American Journal of Roentgenology, **199**(5_supplement):S3-S8 (2012)
[5] Clemens M, Esther M, Marc K, Medical Physics, **38**(2) (2011)
[6] Maas C, Grimmer R, Kachelries M. Dual energy ct material decomposition from inconsistent rays (mdir). Nuclear Science Symposium Conference Record (NSS/MIC), 2009 IEEE, 2009:3446-3452.
[7] Alvarez RE, Macovski A, Physics in medicine and biology, **21**(5):733 (1976)
[8] Tran D, Straka M, Roos J, et al., Academic radiology, **16**(2):160-171 (2009)
[9] Graser A, Johnson TR, Chandarana H, et al., European radiology, **19**(1):13-23 (2009)
[10] F TS, R BC, Marcus H, et al., European journal of radiology, **68**(3) (2008)
[11] Mendler M-H, Bouillet P, Le Sidaner A, et al., Journal of hepatology, **28**(5):785-794 (1998)
[12] Ruzsics B, Lee H, Zwerner PL, et al., European radiology, **18**(11):2414-2424 (2008)
[13] Chae EJ, Song J-W, Seo JB, et al., Radiology, **249**(2):671-681 (2008)
[14] Dong X, Niu T, Zhu L, Medical Physics, **41**(5):051909 (2014)
[15] Niu T, Dong X, Petrongolo M, et al., Medical physics, **41**(4):041901 (2014)
[16] E KP, M AA, A FJ, Technology in Cancer Research and Treatment, **5**(4) (2006)
[17] Barber RF, Sidky EY, Schmidt TG, et al., arXiv:1511.03384v03381 (2015)
[18] Kelcz F, Joseph PM, Hilal SK, Medical physics, **6**(5):418-425 (1979)
[19] Kalender WA, Klotz E, Kostaridou L, IEEE transactions on medical imaging, **7**(3):218-224 (1988)
[20] Warp RJ, J. T. Dobbins, Medical physics, **30**(2):190-198 (2003)
[21] Rutherford R, Pullan B, Isherwoord I, Neuroradiology, **11**(15-21 (1976)
[22] Rutherford RA, Pullan BR, Isherwood I, Neuroradiology, **11**(1):23-28 (1976)
[23] Macovski A, Nishimura D, Doost-Hoseini A, et al., IEEE Trans Med Imaging, **2**(3):122-127 (1983)
[24] Nishimura DG, Macovski A, Brody WR, Medical physics, **11**(3):259-265 (1984)
[25] Zhang R, Thibault J, Bouman C, et al. A model-based iterative algorithm for dual-energy x-ray ct reconstruction. Proceedings of the Second International Conference on Image Formation in X-Ray Computed Tomography. Salt Lake City, Utah, 2012.
[26] Fessler JA, Elbakri IA, Sukovic P, et al., Maximum-likelihood dual-energy tomographic image reconstruction, in *International society for optics and photonics for medical imaging*, edited by 2002), p. 38-49
[27] Candes EJ, Romberg JK, Tao T, IEEE Transactions on Information Theory, **52**(2):489-509 (2006)
[28] Candes EJ, Romberg JK, Tao T, Communications on Pure and Applied Mathematics, **59**(8):1207-1223 (2006)
[29] Sidky EY, Pan X, Physics in Medicine and Biology, **53**(17):4777-4807 (2008)
[30] Feldkamp L, Davis L, Kress J, Journal of Optical Society of America A, **1**(6):612-619 (1984)
[31] Andersen AH, IEEE Transactions on Medical Imaging, **8**(1):50-55 (1989)
[32] Andersen A, Kak A, Ultrasonic Imaging, **6**(1):81-94 (1984)
[33] Glowinski R. *Numerical methods for nonlinear variational problems,* New York: Springer, 1984), p.
[34] Glowinski R, Le Tallec P. *Augmented lagrangian and operatorsplitting methods. In: Nonlinear mechanics,* SIAM, Philadelphia: SIAM Studies in Applied Mathematics., 1989), p.
[35] Bertsekas DP, Tsitsiklis JN. *Parallel and distributed computation: Numerical methods,* NJ: Prentice hall Englewood Cliffs, 1989), p.
[36] Wang L, Cai A, Zhang H, et al., Journal of X-ray Science and Technology, **23**(1):83-99 (2015)
[37] Wang L, Cai A, Zhang H, et al., Computational and Mathematical Methods in Medicine, **2013**((2013)
[38] Cai A, Wang L, Zhang H, et al., Journal of X-Ray Science and Technology, **22**(3):335-349 (2014)
[39] Li J, Niu S, Huang J, et al., PloS one, **10**(10):e0140579 (2015)
[40] Zhang H, Wang L, Yan B, et al., Chinese Physics B, **22**(7):078701 (2013)
[41] Deng W, Yin W. On the global and linear convergence of the generalized alternating direction method of multipliers. 12-14. RICE Computational and Applied Mathematics Technical Report, 2012.
[42] Siddon RL, Medical Physics, **12**(2):252-255 (1985)